Spin-Polarized Tunneling Study on Spin-Momentum Locking in Topological Insulators

Luqiao Liu[1], A. Richardella[2], Ion Garate[3], Yu Zhu[1] N. Samarth[2] and Ching-Tzu Chen[1*]

[1] IBM TJ Watson Research Center, Yorktown Heights, NY, 10598

[2] Department of Physics, The Pennsylvania State University, University Park, PA 16802

[3] Département de Physique and Regroupement Québécois sur les Matériaux de Pointe, Université de Sherbrooke, Sherbrooke, Québec, Canada J1K 2R1

We demonstrate that the charge-spin conversion efficiency of topological insulators (TI) can be experimentally determined by injecting spin-polarized tunneling electrons into a TI. Through a comparative study between bismuth selenide and bismuth antimony telluride, we verified the topological-surface-state origin of the observed giant spin signals. By injecting energetic electrons into bismuth selenide, we further studied the energy dependence of the effective spin polarization at the TI surface. The experimentally verified large spin polarization, as well as our calculations, provides new insights into optimizing TI materials for near room-temperature spintronic applications.

* cchen3@us.ibm.com.

Topological insulators (TI) exhibit coupled spin and momentum orientations in their gapless surface states[1-5]. Recent experiments using TI/ferromagnet bilayer structures suggest that TI can exhibit ultra-high efficiencies in converting electrical charge current into spin accumulation (or vice versa)[6-9], providing great potential for various spintronic applications. To further optimize this effect, it is highly desirable to quantitatively characterize the spin structures of TI, especially in the absence of direct contact between the TI film and ferromagnetic material, as the latter can break time reversal symmetry and change the intrinsic electronic structure of the TI[10]. So far, the mechanism of the observed phenomena – whether they derive mainly from the topological surface states (TSS) [6-8, 11] or from the bulk states[9] – remains largely unclear in TI based spintronic studies. Moreover, large discrepancies both in the magnitude and the temperature dependence of the measured effects still exist in the two currently utilized experimental approaches – injecting a spin current with spin pumping and measuring the induced voltage[8, 9] *versus* applying a charge current and detecting the spin generation[6, 7, 11, 12]. In this work, we demonstrate a new spin-resolved tunneling technique that allows us to quantitatively determine the charge-spin conversion efficiency of topological insulators. By comparing the results from two representative TI materials – bulk-state diluted $Bi_2Se_3$ and surface-state dominated $(Bi_{0.5}Sb_{0.5})_2Te_3$, we confirm the TSS origin of the measured spin signals. The consistent results obtained in the two reversible configurations in our study, along with the observed good thermal stability, reconcile the different experimental results reported earlier. Furthermore, using energetic tunneling electrons, we determine the energy-dependent effective spin polarization of the TSS. With the information gained from these experiments, we infer that further improvement on the charge-spin conversion efficiency can be achieved by tuning the surface carrier ratio and the chemical potential.

The device geometry used in our experiments is illustrated in Fig. 1(a). It is very similar to one we previously used to study the (inverse) spin Hall effect in heavy metals[13]. The devices were patterned on film stacks consisting of TI/MgO(2~3)/$Co_{20}Fe_{60}B_{20}$ (4)/capping layer grown on InP(111)A substrates (units in nanometers), where TI represents 7 quintuple layers (QL) of $Bi_2Se_3$ or 6 QL of $(Bi_{0.5}Sb_{0.5})_2Te_3$.

The stacks were grown in two steps: TI films were first grown by molecular beam epitaxy (MBE)[14] and then transferred through air into the sputtering chamber for oxide barrier and ferromagnetic electrode deposition. In-situ Se capping layers were deposited in the MBE chamber before breaking the vacuum. The capping layer was later removed by 250 °C annealing in the sputtering chamber prior to MgO deposition.

The carrier densities of the $Bi_2Se_3$ and $(Bi_{0.5}Sb_{0.5})_2Te_3$ samples are determined by standard Hall measurements using the van der Pauw method. The 2D carrier density ($n_{2D}$) for $Bi_2Se_3$ is $n_{2D} \sim 2 \times 10^{13}$ $cm^{-2}$ (n-type), equivalent to a 3D carrier density of $\sim 3 \times 10^{19}$ $cm^{-3}$. This value is at least an order of magnitude larger than typical carrier density of a surface-transport dominated TI sample[15], indicating that the Fermi level is located at the bottom of the conduction band, $\sim$ 350meV above the Dirac point (see model calculation in Appendix). For the $(Bi_{0.5}Sb_{0.5})_2Te_3$ film, $n_{2D} \sim 3 \times 10^{12}$ $cm^{-2}$ (n-type), indicating that the Fermi level of the sample is located within the bulk band gap. As is discussed below, this ratio of the total carrier density in the $Bi_2Se_3$ sample to that in $(Bi_{0.5}Sb_{0.5})_2Te_3$ is order-of-magnitude consistent with the spin signal ($R_H$) ratio after taking into account of the difference in the sample sheet resistance.

To achieve optimal spin polarization in the tunnel junction, a 250 °C post-sputtering annealing was applied. The resistance-area (RA) product of the tunnel junctions studied in our experiment ranges from 500 kΩ·µm² to 2 MΩ·µm². Fig. 1(b) shows a typical voltage-dependent junction resistance curve, revealing the tunneling nature of the devices. The highly resistive, crystallized MgO tunneling barrier (see Fig. 1(c) for a typical TEM cross section of the film stacks) ensures a large spin polarization of the tunneling electrons, which helps to overcome the challenge of conductance-mismatch caused reduction in spin signal (or spin injection) that may have plagued some of the reported electrical measurements. The use of a thick barrier also avoids the introduction of magnetic impurities into the TI surface as is in the FM/TI bilayer geometry. To inject spin-polarized electrons into the TSS, we apply AC currents across the tunnel junction between leads 1 and 3. Because of spin-momentum locking, the tunneling electrons gain transverse velocities as they enter the TSS (see Fig. 1(d)), generating a charge accumulation in the orthogonal direction which is detected as a voltage between leads 2 and 4. Compared with previous experiments that used the inverse geometry (injecting current between leads 2 and 4 while measuring the voltage between leads 1 and 3)[11, 12], the tunneling configuration has the advantage of allowing us to investigate the spin texture for states away from the Fermi level by superimposing a DC voltage across the barrier. Moreover, as described below, with our device structure, we can perform measurements in both of the two reversible configurations, providing a self-consistency check of the measured charge-spin conversion. All measurements were carried out at 4K unless otherwise specified.

Fig. 2(a) shows a typical differential resistance vs. magnetic field curve of a $Bi_2Se_3$ sample when the field is swept along the 0° direction (square symbols; see Fig. 1(a) for the definition of field directions). The switching fields in the resistance curve are consistent with the coercive fields of the CoFeB electrode at low temperatures. The transverse voltage changes sign as the magnetization of CoFeB flips direction. First, we notice that the polarity of the sign change is consistent with the helicity of the top TSS, indicating that the observed phenomena come from the Dirac electrons in $Bi_2Se_3$ instead of alternative mechanisms, such as the Rashba state where the opposite spin helicity would dominate[10, 16, 17]. Second, the transverse differential resistance $R_H$ (defined as half of the difference between the saturation values of $dV_{24}/dI_{13}$ at large positive and negative fields) of 60 mΩ is much larger than what we measured in spin Hall metals with similar dimensions (~3 mΩ for Ta and ~0.4 mΩ for Pt). To facilitate the comparison of charge-spin conversion efficiency, we calculate the "effective spin Hall angle" of our $Bi_2Se_3$ sample using Equation (1) in Ref. [13]. The spin-injection induced transverse current in the TSS equals to that generated by a corresponding spin Hall material with a spin Hall angle of ~0.8 if we use the measured sample resistivity of 700 μΩ·cm and take the spin diffusion length along the z direction to be comparable to other spin Hall metals (~ 1 nm)[18, 19]. This value is larger than any known metals.[20-22] Besides 0°, we also measured the transverse resistance with the field applied along the 90° direction (see circles in Fig. 2(a)). In this field orientation, the induced current flows along the *x* direction, and therefore does not generate a voltage between leads 2 and 4. The resistance peaks at low fields in the 90° sweep curve reflect the spontaneous magnetic orientation along the easy axis and the uncompensated magnetic domains during switching.

The extraordinarily large "effective spin Hall angle" observed in $Bi_2Se_3$ samples cannot be explained by the spin Hall effect of the bulk states. By calculating the intrinsic spin Hall conductance using the band structure of the $Bi_2Se_3$, we show that the bulk states can only account for less than 1% of the observed signal (see Fig. 3(a)-(c) and Appendix B). Meanwhile, the large transverse resistance is consistent with the TI surface states origin according to the calculation below. For TSS, the conservation of particle number requires that the injected spin-current density $j_s = \frac{\hbar}{2e}\frac{\eta P_J P_{TI} I}{ab}$ equals to the relaxation rate of the non-equilibrium spins $\frac{\hbar}{2}\frac{\delta n}{\tau_{sf}}$, where $P_J, P_{TI}, \eta$ and $\tau_{sf}$ denote the spin polarization of the junction, spin polarization of the TSS[23-26], the fraction of electrons tunneling into the surface states and the spin flip time, and $a$ and $b$ are the tunnel junction dimensions along the $y$ and $x$ direction. Also, $\delta n = \int \frac{dk}{(2\pi)^2} <k|\sigma_x|k>$ is the spin accumulation at TI surface, with $\sigma_x$ representing the spin Pauli matrix and $|k>$ representing the eigenstates of the TSS at the electrochemical potential. In the presence of spin-momentum locking, the non-equilibrium spins result in a charge-current flow in the direction perpendicular to the spin orientation[27, 28] $j_c = \delta n e v_F$, where $v_F$ is the Fermi velocity. In an open circuit, the induced transverse current $j_c b$ cancels the drift current generated by the voltage, $I_{drift} = \frac{V}{R} = \frac{V}{(a/w)R_\square} = j_c b$. Here $w$ is the width of the TI channel, and $R_\square$ is the total sheet resistance which contains contributions from both the bulk and the surface state carriers. Using the relations above, we get:

$$\frac{dV}{dI} = \eta P_J P_{TI} R_\square \frac{v_F \tau_{sf}}{w} \approx \eta P_J P_{TI} R_\square \frac{l}{w}. \tag{1}$$

In deriving Eq. (1), we note that $\tau_{sf}$ roughly equals the momentum relaxation time[28]; hence the mean free path $l \approx v_F \tau_{sf}$. Using the measured values of $P_J \approx 0.5$, $R_\square \approx 1\ k\Omega$ and $w = 8\ \mu m$ for our device and the reported values of $l \sim 30 - 130$ nm [15, 29] and $P_{TI} \sim 0.4$ [26] for $Bi_2Se_3$ TSS, we estimate that the portion of electrons which tunnel into the surface states $\eta$ is roughly on the order of magnitude of 0.01 – 0.1. This $\eta$ value is consistent with the ratio of the TSS carrier density to the total carrier density measured in our $Bi_2Se_3$ sample (see Fig. 3(b) and Appendix).

Besides detecting the transverse voltage induced by the spin-polarized tunneling electrons, we have also conducted the measurement using the inverse geometry, i.e., applying the charge current between leads 2 and 4 and measuring the transverse voltage between leads 1 and 3, as shown in Fig. 2(b). This measurement configuration is similar to previous studies in Ref. [11] and [12]. The junction voltage measured in this geometry comes from the electrochemical potential difference across the tunneling barrier (see the inset of Fig. 2(b)). Applying the Einstein relation to the 2D TSS, we can derive the equation for the junction voltage, $V_{13} = \eta P_J P_{TI} I_{24} R_\square l/w$ [23, 30], the same as Equation (1). Comparing Fig. 2(a) and (b), we see that the signal magnitudes are almost the same. This consistency proves the equivalence of the two experimental configurations.

In previous studies on the electrical detection of spin-momentum locking, strong temperature dependence has been reported[11, 12], which casts doubt on the robustness of the observed charge-spin conversion efficiency for applications at elevated temperatures. On the other hand, efficient room-temperature spin-torque generation[7] and spin-pumping induced spin-charge conversion[31] have been reported. The large discrepancy in the temperature dependence invites the question about whether the aforementioned experiments probe the same physics or not. To address this issue, we check the thermal stability of the measured tunneling spin signals. We vary the $Bi_2Se_3$ sample temperature from 4 K to 200 K and, as illustrated in Fig. 4(a), observe very small changes in the transverse resistance $R_H$, a strong indication of the robustness of the TSS spin-momentum coupling. This result contrasts the reports in Ref. [11] and [12] where the spin-momentum locking induced signal drops significantly with rising temperatures. One possible explanation is that the well-crystallized MgO barrier in our devices ensures high junction spin polarization $P_J$ at elevated temperatures. The consistency between our results and the previous room-temperature spin-torque and spin-pumping experiments suggests that the observed effects share similar origins.

Further information about the $Bi_2Se_3$ samples can be extracted by applying finite voltages across the tunnel junction and measuring the changes in the transverse resistance $R_H$ in the tunneling configuration. We plot a series of transverse resistance curves under different DC biases $V_{DC}$ in Fig. 4(b). A compilation of the results for four representative devices is shown in Fig. 4(c). All the devices exhibit similar asymmetric bias dependence. We note that, in the voltage-dependence experiments, the bias voltage across the tunnel junction only varies the final states that we probe. The Fermi level of the TI sample remains the same. Therefore, the backflow drift current experiences the same sheet resistance $R_\square$ (see Eq. (1)). Variations in $R_H$ thus reflect only the variations in $\eta$, $P_J$, $P_{TI}$ and $l$ with energy.

Under positive $V_{DC}$, a steep drop in $R_H$ is observed above 150 mV. In comparison, under negative $V_{DC}$, $R_H$ begins to decrease gradually at low bias. Here we adopt the convention that under positive bias, electrons tunnel from CoFeB into the unoccupied states in $Bi_2Se_3$ (see Fig. 1(a)). From the ARPES experiment on similar samples (Ref. [26]), the location of the Fermi surface is determined to be pinned to the bottom of the conduction band and the Dirac point is about 300 – 350 meV below the Fermi level. Besides, we have also calculated the sample Fermi level position using the carrier density of our as-grown $Bi_2Se_3$ films (~3 ×$10^{19}$ cm$^{-3}$) in a 4-band model Hamiltonian[32] and get consistent results (see Appendix A and Fig. 3(b)). According to first-principle calculations[25, 33], the TSS in $Bi_2Se_3$ merges with the bulk band at ≥ 100 meV above the conduction band edge. Electrons tunneling into the surface states with higher energy quickly gain bulk characteristics in that their wave function extends across the entire thickness of the film, connecting the top and bottom surfaces with opposite helicities. The rapid drop of $R_H$ at $V_{DC}$ > +150 mV thus reflects the cancelation of spin polarization $P_{TI}$ in the TSS.

Under negative $V_{DC}$, the tunneling spectroscopy probes the occupied states below the Fermi level into the bulk gap. Since the mean-free path $l$ [34, 35] and junction $P_J$ [13] vary only slightly within the voltage range studied, the $R_H$ vs. $V_{DC}$ variation predominantly reflects the energy dependence of the effective spin polarization in our samples $\eta P_{TI}$ (see Eq. (1)), providing a quantitative measure of the TSS spin polarization accessible in real devices. Because the density of states of the TSS decreases linearly as we probe deeper into the bulk gap (the 2D density of states $N \propto k_F$), in the presence of the many mid-gap defect states originated from the Se vacancies and other disorder, the probability ($\eta$) of electrons tunneling into the TSS also drops quasi-linearly. This is manifested in the quasi-linear decrease in the observed tunneling spin signals $R_H$ (and thus $\eta P_{TI}$) with negative $V_{DC}$ (Fig. 4(c)). The energy dependence of $\eta P_{TI}$ is also consistent with the ARPES results on similar $Bi_2Se_3$ films (with thickness $\geq$ 6QL) that, upon approaching the Dirac point, the TSS spin polarization $P_{TI}$ only decreases slightly until the states are very close to the Dirac point[26].

Besides the model calculations and the voltage dependence data of $Bi_2Se_3$, a comparative study of the $(Bi_{0.5}Sb_{0.5})_2Te_3$ sample provides another piece of evidence that the charge-spin conversion mostly originates from the TSS instead of the bulk states. In our MBE-grown $(Bi_{0.5}Sb_{0.5})_2Te_3$ films, the 2D carrier density is ~3 ×10$^{12}$ cm$^{-2}$, indicative of TSS dominated density of states at the Fermi level, consistent with previous photoemission experiment[36]. Correspondingly, compared with $Bi_2Se_3$, the $(Bi_{0.5}Sb_{0.5})_2Te_3$ sample should exhibit an enhanced transverse resistance $R_H$ if the signal mainly comes from the TSS and a diminished $R_H$ if it mostly comes from the bulk. Fig. 5 plots the $dV_{24}/dI_{13}$ vs. $H$ data in the tunneling configuration, with the field applied along both 0° and 90° directions. As in $Bi_2Se_3$, the switching curves are consistent with the symmetry and helicity of the TSS spin-momentum locking in TI. However, the observed signal, $R_H \sim 10\ \Omega$, increases by more than 2 orders of magnitude. After taking into account of material resistivity contribution, the enhancement in $R_H$ leads to an "effective spin Hall angle" of ~ 20 ± 5, making $(Bi_{0.5}Sb_{0.5})_2Te_3$ one of the most efficient materials for converting electron charge into spin. Using $P_J \approx 0.5$, $R_\square \approx 6\ k\Omega$, $w = 8\ \mu m$, $l \sim 150 \pm 50$ nm [37, 38] and $P_{TI} \sim 0.2 - 0.45$ [5, 39], we can calculate the probability of electrons tunneling into the TSS $\eta \sim 0.6 \pm 0.2$ from equation (1). This large $\eta$ implies that a majority of the injection carriers tunnel into the TSS in the (Bi$_{0.5}$Sb$_{0.5}$)$_2$Te$_3$ film, consistent with the measured carrier densities.

As the experiment on (Bi$_{0.5}$Sb$_{0.5}$)$_2$Te$_3$ samples reveals a large efficiency improvement when the surface state dominates at the Fermi level, a quick calculation of $R_H$ in the case that only the TSS contributes to the electrical transport would provide further insight into optimizing charge-spin conversion. Under the assumption that $\eta = 1$ and $R_\square = \frac{1}{\sigma_{SS}}$, Equation (1) can be re-written as:

$$\frac{dV}{dI} = \frac{2h}{e^2}\frac{P_J P_{TI}}{k_F w} \approx 5.2\,\Omega \cdot P_J P_{TI} (\frac{\text{Å}^{-1}}{k_F})(\frac{\mu m}{w}). \tag{2}$$

Here we utilize the TSS conductivity, $\sigma_{SS} = \frac{e^2}{2h} k_F l$. Equation (2) gives the quantum limit of the charge-current induced spin accumulation or the spin-injection induced electric signal in TI samples. According to Ref. [26] and the analysis in the previous paragraphs, TSS spin polarization $P_{TI}$ does not decrease much until the Fermi level comes very close to the Dirac point for film thickness ≥ 6QL. Under this condition, the relationship of $R_H \propto \frac{1}{k_F}$ implies that the charge-spin conversion efficiency peaks when the Fermi level approaches the Dirac point. In the configuration of Fig. 2(b), a small applied charge current can generate a large effective spin quasi-electrochemical potential $\delta\mu$ at the TI surface. For a given interface formed by TI and another material with a fixed spin mixing conductance, this will lead to very efficient spin-current injection. Therefore, Equation (2) predicts that maximum charge-spin conversion can be realized by tuning the Fermi level of TI close to the Dirac point, which can be achieved by chemical doping or back gating.

In summary, using a tunnel junction based spin-polarized tunneling technique, we quantitatively determine the charge-spin conversion efficiency of two representative topological insulators. The robust MgO tunnel barrier used in our experiment enables an accurate measurement by eliminating the reduction in spin signals from pinholes and conductance mismatch. By comparing the results of Bi$_2$Se$_3$ and (Bi$_{0.5}$Sb$_{0.5}$)$_2$Te$_3$, we verify the TSS origin of the observed spin signals. The exceptionally large charge-spin conversion efficiency and the temperature stability suggest that the protected TSS in topological insulators can provide a very promising platform for near room temperature spintronic devices.

**Acknowledgements**

We thank Hsin Lin, Peng Wei, Michael Flatté, Cuizu Chang, Jonathan Sun, Guang-Yu Guo, and Joon Sue Lee for illuminating discussions. We are grateful for the technical assistance of Jemima Gonsalves, Florian Lemaitre, and Andrew Argouin. The work at IBM and Penn State is supported by the DARPA MESO program (N66001-11-1-4110). N.S. and A.R. also acknowledge partial support from ONR N000141210117. I.G. acknowledges funding support from Université de Sherbrooke and from the National Science and Engineering Research Council of Canada, and computer support from Calcul Québec and Compute Canada.

**APPENDIX**

In the Appendix section, we present the theoretical modeling of the experiment, concentrating on $Bi_2Se_3$ samples. The theory demonstrates that the transverse differential resistance $R_H$ measured in the main text originates predominantly from the spin-momentum locking of the topological surface states, via the spin-galvanic effect. The bulk state intrinsic spin Hall effect contribution is less than 1%. The outline of the Appendix is as follows:

*A. Model Hamiltonians in $Bi_2Se_3$.*

*B. Theoretical modeling of the measured signal $R_H$:*

*B-1. Contribution from bulk band spin Hall effect in $Bi_2Se_3$.*

*B-2. Contribution from the spin-galvanic effect in $Bi_2Se_3$.*

*B-3. Total **measured transverse** signal $R_H$.*

## APPENDIX A. MODEL HAMILTONIANS IN $Bi_2Se_3$

### A-1. Bulk states

The model Hamiltonian describing the electronic structure of bulk $Bi_2Se_3$ near the bottom of the conduction band and the top of the valence band is given by[32, 40]:

$$h_B(\boldsymbol{k}) = \epsilon_{\boldsymbol{k}} + \boldsymbol{d}_{\boldsymbol{k}} \cdot \boldsymbol{\sigma}\tau^x + M_{\boldsymbol{k}}\tau^z, \tag{A.1}$$

where $\epsilon_{\boldsymbol{k}} = \gamma_x(k_x^2 + k_y^2) + \gamma_z k_z^2$, $\boldsymbol{d}_{\boldsymbol{k}} = (v_x k_x, v_y k_y, v_z k_z)$, $M_{\boldsymbol{k}} = M - t_x(k_x^2 + k_y^2) - t_z k_z^2$. Below, we will extrapolate Eq. (A.1) for momenta that lie far from the center of the Brillouin zone, by resorting to a simplified (tetragonal) lattice model with lattice constants that match those of $Bi_2Se_3$ ($a_x$ = 4 Å, $a_z$ = 30 Å).

A fit to *ab-initio* band structure calculation yields [32] $M = -0.28$ eV, $t_z = -10$ eV Å$^2$, $t_x = -56$ eV Å$^2$, $v_z = 2.2$ eV Å, $v_x = 4.1$ eV Å, $\gamma_z = 1.3$ eV Å$^2$ and $\gamma_x = 19.6$ eV Å$^2$. Here, the *x* and *y* directions are parallel to the quintuple layers, whereas the *z* direction is perpendicular to them. Following the convention of Fig. 1, the *x* direction points from terminal 1 to terminal 3, while the *y* direction points from terminal 2 to terminal 4. Also, $\boldsymbol{\sigma}$ and $\boldsymbol{\tau}$ represent Pauli matrices: $\sigma^z \in \{\uparrow, \downarrow\}$ denotes the projections of the *z*-component of the total angular momentum, whereas $\tau^z \in \{P1, P2\}$ labels atomic orbitals of opposite parity (with respect to spatial inversion about the inner Se atom in the quintuple layer).

The diagonalization of Eq. (A.1) produces eigenstates $|\boldsymbol{k}n; B\rangle$ and eigenvalues $E_{\boldsymbol{k}n}^{(B)}$,

$$h_B(\boldsymbol{k})|\boldsymbol{k}n; B\rangle = E_{\boldsymbol{k}n}^{(B)}|\boldsymbol{k}n; B\rangle, \tag{A.2}$$

where $n \in \{1,2,3,4\}$ is the band label.

### A-2. Surface states

The model Hamiltonian describing the topological surface states at a surface of $Bi_2Se_3$, which is parallel to the quintuple layers, is given by[40]:

$$h_S(\boldsymbol{k}) = \epsilon_0 + A_1(\sigma^x k_y - \sigma^y k_x) + A_2(k_x^2 + k_y^2) + A_3(k_+^3 + k_-^3)\sigma^z, \tag{A.3}$$

where $\sigma^i$ are spin Pauli matrices, $k_\pm = k_x \pm ik_y$, $A_1 = 3.33$ eV Å is the Dirac velocity, $A_2 = 23.73$ eV Å$^2$, $A_3 = 25.3$ eV Å$^3$ is the coefficient for hexagonal warping and $\epsilon_0 = 0.034$ eV. In the calculations below, we take a momentum cutoff of $k_c = 0.075$ Å$^{-1}$.

The diagonalization of Eq. (A.3) produces eigenstates $|kn;S\rangle$ and eigenvalues $E_{kn}^{(S)}$,

$$h_S(\mathbf{k})|kn;S\rangle = E_{kn}^{(S)}|kn;S\rangle, \tag{A.4}$$

where $n \in \{1,2\}$ is the surface band label.

## APPENDIX B. THEORETICAL MODELING OF THE MEASURED SIGNAL R$_H$

### B-1. Contribution from bulk band spin Hall effect in Bi$_2$Se$_3$

An obvious mechanism contributing to R$_H$ in our experiment is the inverse Spin Hall effect, whereby a longitudinal spin current gives rise to a transverse electric field. The starting point to quantify the inverse spin Hall effect is to compute the spin Hall conductivity. In this section, we calculate the intrinsic (i.e. band-structure) spin Hall conductivities of the bulk bands, and conclude that the corresponding inverse spin Hall effects cannot explain our data.

The intrinsic contribution to the bulk-state spin Hall conductivity can be written as (see e.g. Ref.[41]

$$\sigma_{ij;B}^\lambda = e^2 \int \frac{d^3k}{(2\pi)^3} \sum_n f_{kn}^{(B)} \sum_{n'\neq n} \frac{Im[\langle kn;B|J_{i;B}^\lambda|kn';B\rangle\langle kn';B|v_j^{(B)}|kn;B\rangle]}{(E_{kn}^{(B)} - E_{kn'}^{(B)})^2}, \tag{A.5}$$

where $f_{kn}^{(B)}$ is the Fermi occupation factor for the bulk state $|kn;B\rangle$, $v_i^{(B)} = \partial h_B(\mathbf{k})/(\hbar\partial k_i)$ is the velocity operator and

$$J_{i;B}^\lambda = \hbar\{\sigma^\lambda, v_i^{(B)}\} \tag{A.6}$$

is the spin current operator in the bulk. Physically, $\sigma_{ij;B}^\lambda$ describes the $\lambda$-spin current flowing in the bulk along the $i$ direction, in response to an electric field applied along $j$. Throughout this section, we take the convention that the spin current has the same physical dimensions as the charge current.

The only nonzero components of $\sigma^{\lambda}_{ij;B}$ are those for which $i$, $j$ and $k$ are different from one another. In our experiment, there is an $x$-spin current that is injected into Bi$_2$Se$_3$ from the tunnel junction. The current-carrying electrons are spin-polarized by the ferromagnetic electrode adjacent to the tunnel barrier. This non-equilibrium spin polarization diffuses into the bulk of Bi$_2$Se$_3$, which then results in an $x$-spin current flowing along $z$. Because $\sigma^{x}_{zy;B} \neq 0$ (cf. Fig. 6), this spin current creates an electric field along the y direction, which is then measured as a Hall voltage. At first glance, this suggests that the bulk inverse spin Hall effect might be a serious candidate to explain our data. However, as we show next, this is not the case. The inverse spin-Hall contribution from the bulk states to the Hall resistance is given by[13]:

$$R_H^{bulk} = \frac{\sigma^{x}_{zy;B}}{(\sigma_B^c)^2} \frac{P}{w} \frac{\lambda_s}{t} \tanh\left(\frac{t}{2\lambda_s}\right), \tag{A.7}$$

where $t \simeq 7$ nm is the thickness of the topological insulator film, $w \simeq 8$ μm is the width of the tunneling device along the x direction, $\lambda_s$ is the bulk spin diffusion length along the z direction ($\lambda_s \sim 1$ nm), $\sigma_B^c$ is the longitudinal (in-plane) charge conductivity, and $P \in [0,1]$ is the effective spin polarization of the injected current (in the notation of the main text, $P \equiv P_J P_{TI}$). A nonzero transverse signal requires the breaking of time reversal symmetry, i.e. $P \neq 0$. The in-plane charge conductivity of the bulk states, $\sigma_B^c$, can be calculated as

$$\sigma_B^c = \frac{e^2 \hbar}{2\pi} Re \sum_{nn'} \int \frac{d^3k}{(2\pi)^3} \left|\langle \mathbf{k}n; B | v_x^{(B)} | \mathbf{k}n'; B \rangle\right|^2 [G_{n;B}^R(\mathbf{k}) G_{n';B}^A(\mathbf{k}) - G_{n;B}^R(\mathbf{k}) G_{n';B}^R(\mathbf{k})], \tag{A.8}$$

where $G_{n;B}^{R(A)}(\mathbf{k}) = [\epsilon_F - E_{\mathbf{k}n}^{(B)} + (-)i\hbar/2\tau_B]^{-1}$ is the zero-frequency retarded (advanced) Green's function and $\tau_B$ is the momentum scattering time in the bulk.

Equation (A.7) is applicable only in the metallic bulk regime, where $\sigma_B^c \gg |\sigma^{x}_{zy}|$. This condition is well satisfied in our Bi$_2$Se$_3$ films. In our experiment, the Fermi energy lies close to the bulk conduction band edge. In such situation, the difference between momentum- and transport-scattering times is small[42] and accordingly disorder vertex corrections have been neglected in Eq. (A.8). Incidentally, we note that the in-plane conductivity is much larger than the out-of-plane conductivity. The origin of this anisotropy resides on the layered crystal structure of $Bi_2Se_3$, which makes interlayer transport weaker than in-plane transport.

In Fig. 7, we plot Eq. (A.7) as a function of the Fermi energy. In this plot, we have chosen $\tau_B$ so that the bulk mean free path is $l_B \sim 10$ nm. In addition, following the main text we have taken $P \sim 0.2$. Clearly, the theoretical result is about two orders of magnitude too small to account for our data. The reason for this outcome is mainly that the bulk spin Hall angle $\theta_{SH}^{bulk} \equiv \sigma_{zy;B}^x / \sigma_B^c$ is made particularly small by the crystalline anisotropy because the spin current flows perpendicular to the quintuple layers whereas the charge current backflow producing the Hall voltage flows parallel to them. On this basis, we conclude that the inverse spin Hall effect coming from the bulk states is unimportant in our experiment.

**B-2. Contribution from the spin-galvanic effect in $Bi_2Se_3$**

In the preceding section, we have shown theoretically that the inverse spin Hall effect from bulk state cannot account for the experimentally observed Hall resistance. In this section, we discuss a related (but distinct) mechanism that emerges from the spin-momentum locking of the surface states and appears to explain our data.

As discussed in the main text, injecting a non-equilibrium spin density on the surface is akin to creating an imbalance between right- and left-moving surface electrons. This phenomenon is the inverse of the so-called Edelstein effect, wherein a charge current gets spin polarized due to spin-orbit coupling (see e.g. Ref. [43]). In our context, the spin-galvanic effect arises due to the following properties of the topological surface states: (i) spin-momentum locking, and (ii) and odd (one, in case of $Bi_2Se_3$) number of Fermi crossings in half of the surface-projected Brillouin zone.

In the main text, the spin-galvanic contribution from the surface states to the measured signal has been claimed to be

$$R_H^{surf} = R_\square \frac{l_S P}{w}, \tag{A.9}$$

where $l_S$ is the surface mean free path. Let us derive Eq. (A.9).

We begin with the expression for the expectation value (per unit area) of the $x$-spin operator. Anticipating the fact that the Edelstein effect is present only on the surface states, we write

$$\frac{\langle \sigma^x \rangle}{A} = \int \frac{d^2k}{(2\pi)^2} \sum_n f_{kn}^{(S)} \langle kn; S | \sigma^x | kn; S \rangle, \tag{A.10}$$

where $n$ labels surface bands. In the main text, we have referred to as $\langle \sigma^x \rangle / A$ as the spin accumulation $\delta n$. In equilibrium, $\langle \sigma^x \rangle = 0$. In presence of a weak transport current, we have

$$f_{kn}^{(S)} \simeq f_{kn}^{(S);0} + e\boldsymbol{v}_{kn}^{(S)} \cdot \boldsymbol{E} \tau_{kn;S} \frac{\partial f_{kn}^{(S);0}}{\partial E_{kn}^{(S)}}, \tag{A.11}$$

where $f_{kn}^{(S);0}$ stands for the Fermi-Dirac distribution, $\boldsymbol{v}_{kn}^{(S)} = \partial E_{kn}^{(S)}/(\hbar \partial \boldsymbol{k})$ is the group velocity of surface electrons, $\boldsymbol{E} = (E_x, E_y)$ is the transport electric field, and $\tau_{kn;S}$ is the momentum scattering time associated to Bloch state $|kn; S\rangle$. The second term in Eq. (A.11) leads to a non-equilibrium spin polarization.

However, by neglecting the electric-field-induced changes to the spin matrix elements, we are limiting ourselves to the intraband contribution to the Edelstein effect[44]. We have verified that, for highly conducting surface states (i.e. Fermi energy not very close to the surface Dirac point), the neglect of the interband contribution does not incur a significant error. Moreover, for analytical simplicity, let us ignore quadratic and cubic terms in Eq. (A.3) and let us assume that $\tau_{kn;S} = \tau_S$ for any $\boldsymbol{k}$ and $n$. We have verified numerically that these approximations are good in our case. Then, substituting Eq. (A.11) in Eq. (A.10), it is easy to obtain

$$\frac{\langle \sigma^x \rangle}{A} = -\frac{e}{4\pi\hbar} E_y \tau_S k_F, \tag{A.12}$$

where $k_F$ is the Fermi momentum of the surface states. Equation (A.12) is an example of the well-known Edelstein effect. We can now turn the tables around and obtain an expression for the electric field induced by a non-equilibrium $x$-spin density:

$$E_y = -\frac{4\pi\hbar}{e} \frac{1}{k_F \tau_S} \frac{\langle \sigma^x \rangle}{A}. \tag{A.13}$$

This equation describes the spin-galvanic (or "inverse-Edelstein") effect. From Drude's formula for the charge conductivity, Eq. (A.13) can be rewritten as

$$E_y = -\frac{ev_F}{\sigma_S^C}\frac{\langle\sigma^x\rangle}{A} = -\frac{ev_F}{\sigma_S^C}\frac{PI_{13}\tau_{sf}/e}{ab}, \tag{A.14}$$

where $I_{13}$ is the charge current injected onto the topological surface state from the tunnel junction, $\tau_{sf}$ is the spin relaxation time, and $ab$ is the area of the tunnel junction [$a(b)$ being the linear dimension of the tunnel junction along the $y(x)$ direction]. Due to spin-momentum locking of the topological surface states, $\tau_{sf} \sim \tau_S$ and hence $v_F\tau_{sf}$ may be identified with the surface mean free path $l_S$.

The electric field in Eq. (A.14) produces a surface current along $y$ (i.e. between terminals 2 and 4). Assuming that the induced current density is spatially uniform, the total current induced by the spin-galvanic effect is

$$\sigma_S^C E_y b = l_S P I_{13}/a. \tag{A.15}$$

Under open circuit conditions along the $y$ direction, this current must be cancelled by a backflow current that covers the entire cross section of the transport channel (of area $wt$) and produces a voltage

$$V_{24} = \left(\rho\frac{a}{wt}\right)\sigma_S^C E_y b = \rho\frac{l_S}{wt}PI_{13}. \tag{A.16}$$

Here, $\rho = 1/(2\sigma_S^C/t + \sigma_B^C)$ is the longitudinal resistivity of the topological insulator film (for simplicity, we assume that the two surfaces have the same conductivity). Defining the square resistance of the film as $R_\square$, we arrive at

$$\frac{V_{24}}{I_{13}} = R_\square \frac{l_S}{w} P, \tag{A.17}$$

which is the desired Eq. (A.9).

The longitudinal charge conductivity of the surface states appearing in the expression for $R_\square$ can be calculated from

$$\sigma_S^C = \frac{e^2\hbar}{2\pi}Re\sum_{nn'}\int\frac{d^2k}{(2\pi)^2}\left|\langle \bm{k}n;S|v_x^{(S)}|\bm{k}n';S\rangle\right|^2[G_{n;S}^R(\bm{k})G_{n';S}^A(\bm{k}) - G_{n;S}^R(\bm{k})G_{n';S}^R(\bm{k})], \tag{A.18}$$

where $G_{n;S}^{R(A)}(\bm{k}) = [\epsilon_F - E_{\bm{k}n}^{(S)} + (-)i\hbar/2\tau_S]^{-1}$ is the zero-frequency retarded (advanced) Green's function and $\tau_S$ is the momentum scattering rate on the surface (which in general need not be the same as that in the bulk). In Eq. (A.18), we have ignored disorder vertex corrections. This results in an error of about a factor of two because, in absence of hexagonal warping, the transport scattering time is twice the momentum scattering time[42]. However, such an error will turn out to be unimportant in our estimates below, because we will take $\tau_S$ to be a phenomenological parameter.

In Fig. 8, we plot Eq. (A.9) as a function of the Fermi energy. In this plot, we have chosen $\tau_S$ so that $l_S \sim 20$ nm. This choice is motivated by the fact that we see no evidence of quantum oscillations at high magnetic fields, which implies a surface mean free path that is $\leq 30$ nm. In addition, following the main text we have taken $P \sim 0.2$. With this in mind, Fig. 8 shows that Eq. (A.9) is in good quantitative agreement with the experiment, once we account for the fact that only a given fraction $\eta < 1$ of the current injected from the tunnel junction flows through the surface states (the rest flows in the bulk states). The fraction $\eta$ is calculated in Section B-3. (cf. Eq. (A.20)).

**B-3. Total measured transverse signal $R_H$**

Thus far we have discussed the separate bulk and surface contributions to $R_H$. In this section, we shall combine the preceding results. To simplify the discussion, we suppose that the bulk and surface states in $Bi_2Se_3$ constitute parallel conduction channels. This means that we are neglecting bulk-surface coupling, which is a good approximation if the surface-to-bulk scattering length exceeds the bulk and surface spin diffusion lengths. Then, we have

$$R_H = \eta R_H^{surf} + (1-\eta) R_H^{Bulk}, \quad (A.19)$$

where $R_H^{surf}$ is the Hall resistance coming from the surface states, $R_H^{Bulk}$ is the spin Hall resistance coming from the bulk state and $\eta$ is the fraction of the injected current that flows on the surface. In our tunneling experiment,

$$\eta = \frac{\nu_{surf}}{\nu_{surf} + \nu_{bulk} t}, \quad (A.20)$$

where $\nu_{surf}$ and $\nu_{bulk}$ are the surface and bulk density of states at the electrochemical potential, respectively. Only the top surface is included in $\nu_{surf}$. For the experimentally relevant parameter space, the calculated fraction of the charge current flowing on the surface is $\sim 10 - 20\%$.

For a reality check of Eq. (A.19), let us consider some special cases. If the charge conductivity of the bulk states vanishes (say, because the Fermi level is deep inside the bulk gap), all the injected current flows on the surface and accordingly $R_H \simeq R_H^{surf}$. If the film is thick enough, such that $\nu_{surf} \ll \nu_{bulk} t$, most of the current travels in the bulk and accordingly $R_H \simeq R_H^{bulk}$. If the spin Hall angle of the bulk states vanishes (i.e. if $R_H^{bulk} = 0$), then the measured Hall resistance comes fully from the surface (although not all the injected charge current participates in it because part of it travels into the bulk). These are all reasonable results.

In Fig. 3(c), which is simply a combination of Figs. 7 and 8, we show that the theoretically calculated contribution to the Hall resistance. We find that $R_H \simeq 60\ m\Omega$ for $n \sim 3 \times 10^{19}\ cm^{-3}$, which is in good agreement with experiment. The figure also provides clear theoretical support to the claim that the observed signal originates from the surface states.

We conclude by mentioning the limitations of our theory. First, it is based on the 4-band $\mathbf{k} \cdot \mathbf{p}$ approximation, which gives a correct account of the low-energy electronic structure only close to the Γ point. Second, it neglects electron quantization effects, which may play some role in the bulk states of a 7 nm thick film[45]. At any rate, no evidence of van Hove singularities has been found in experiment, suggesting that the mean free path along the $z$ direction is not large compared to the film thickness. Despite the limitations of our theory, it succeeds at predicting values of $R_H$ that are reasonably close to the measured ones, at the carrier density and longitudinal resistivity corresponding to the measured sample values. It also predicts that the measured signal originates predominantly from the spin-galvanic effect happening on the topological surface state in contact with the tunnel junction. The validity of this statement is likely immune to the limitations of our theory because it rests on the fact that, in layered topological insulators, in-plane conductivity is large compared to interlayer conductivity.

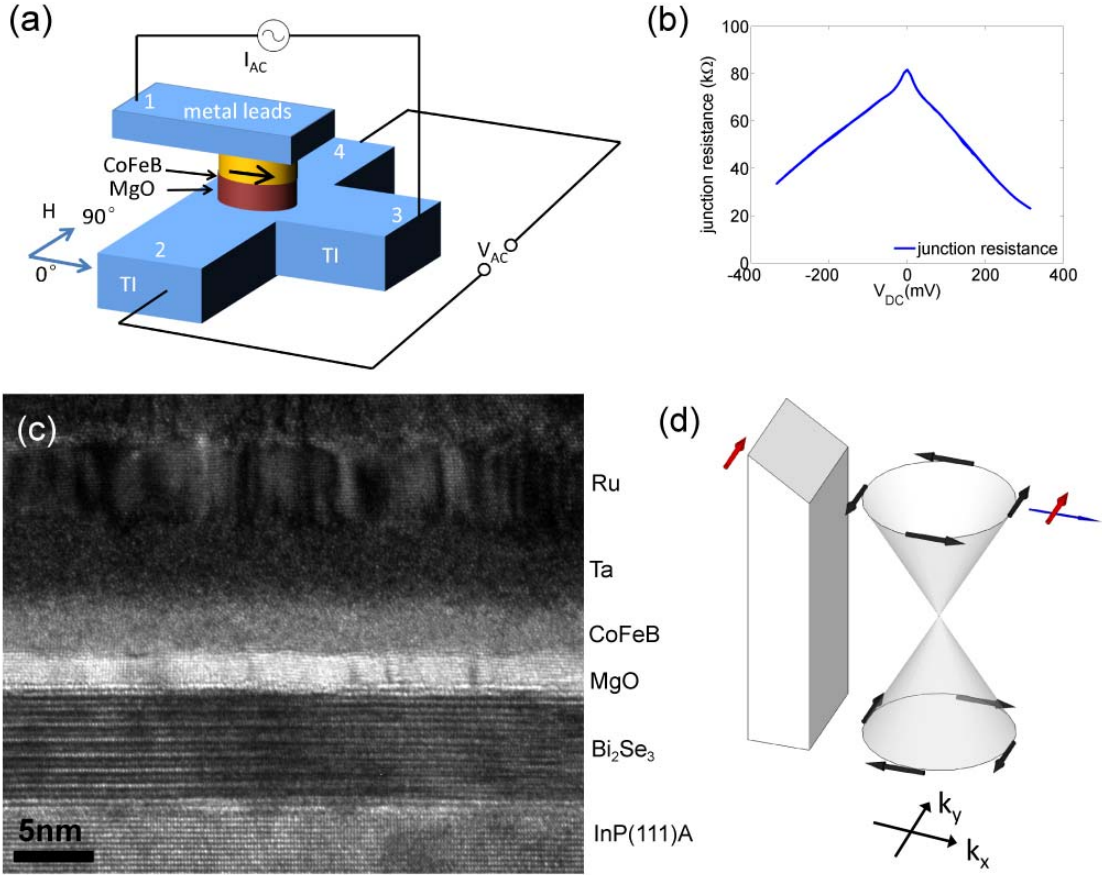

Figure 1. (a) Schematic of the spin-polarized tunnel junction device used in our experiment. The junction area varies from 2 × 2 to 10 × 5 μm$^2$; the width of TI channels is fixed at 8 μm. (b) Voltage-dependent tunnel junction resistance $dV_{13}/dI_{13}$ of a typical device with area 6 × 4 μm$^2$ and RA ~ 1.9 MΩ·μm$^2$. The lead resistance has been subtracted. (c) Transmission electron microscope image of an annealed film stack. (d) Schematic illustration showing the generation of a charge current when spin-polarized tunneling electrons are injected into the TSS.

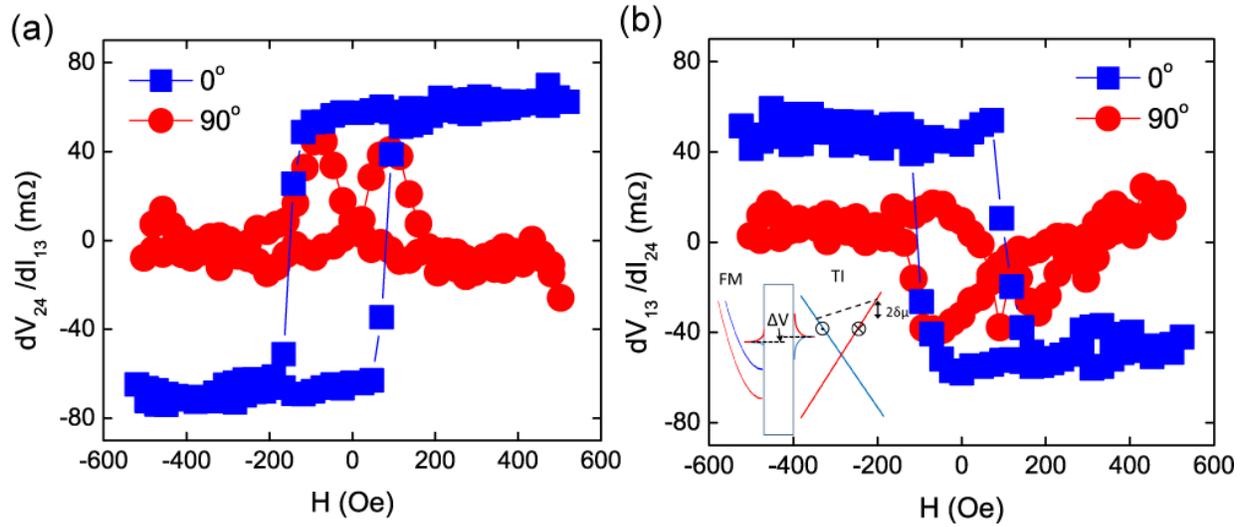

Figure 2. (a) Field-dependent transverse differential resistance $dV_{24}/dI_{13}$ of $Bi_2Se_3$ in the tunneling configuration. The junction area is 10×3 μm$^2$. (b) Field-dependent transverse resistance $dV_{13}/dI_{24}$ of the same device in the inverse configuration. The magnetic field is swept along 0° (square) and 90° (circle) respectively. Inset of (b): Illustration of the formation of transverse voltage across the tunnel barrier when a charge current is applied along the TSS. The spin-dependent electrochemical potential δμ is caused by the current-induced spin accumulation.

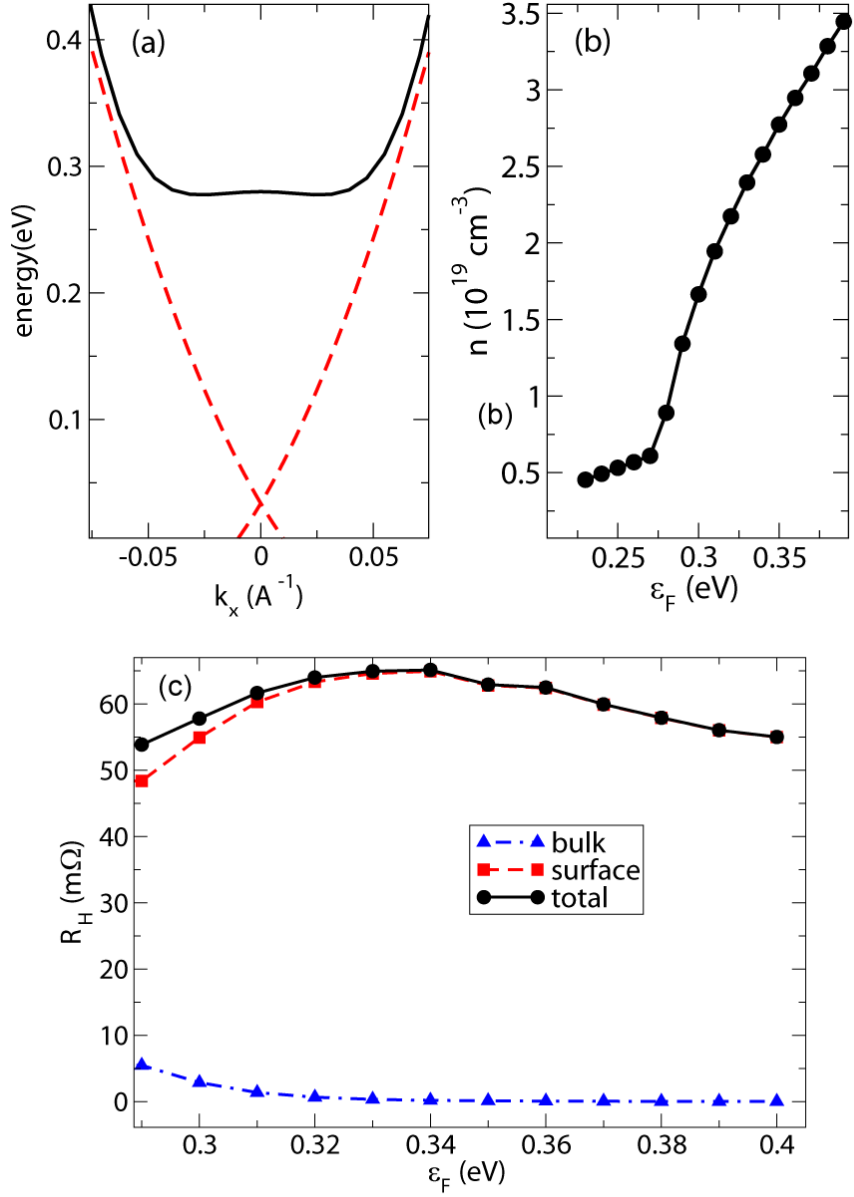

Figure 3 (a) Energy dispersion of the bulk conduction band edge (black line) and the topological surface states (red dashed lines) near the Γ point of $Bi_2Se_3$. The bulk conduction band edge is situated at 0.28 eV. (b) Calculated carrier density as a function of the Fermi energy. In the calculation of the density, we have used $n = n_{surf}/t + n_{bulk}$, where $n_{surf}$ ($n_{bulk}$) is the surface (bulk) carrier density and t = 7 nm is the film thickness. A comparison with the experimentally measured carrier density indicates that $\varepsilon_F \simeq 0.35$ eV for $Bi_2Se_3$ sample. (c) Calculated spin signal $R_H$, displaying the surface and bulk contributions as well as the total. The calculation details are given in Appendix A and B.

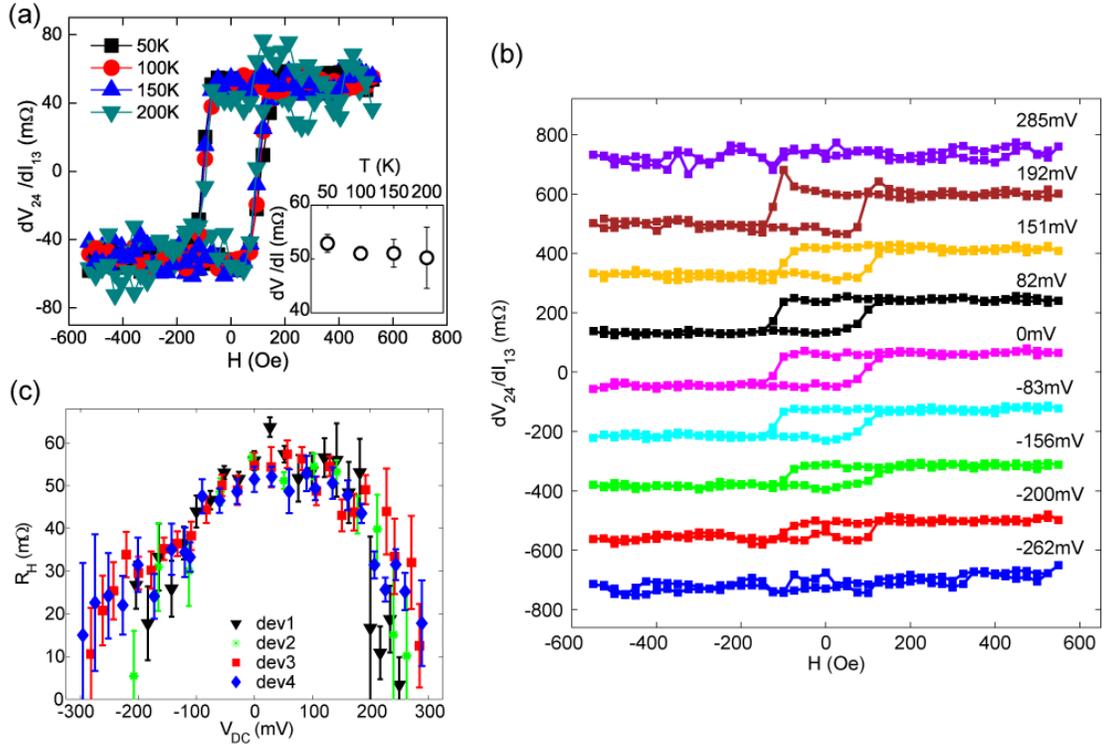

Figure 4 (a) Field-dependent transverse resistance of $Bi_2Se_3$ at zero bias measured under different temperatures. The effect of the temperature on the magnitude of $R_H$ is summarized in the inset. (b) Transverse differential resistance $dV_{24}/dI_{13}$ vs. applied magnetic field $H$ under different DC bias voltages $V_{DC}$ across the tunnel junction of CoFeB|MgO|$Bi_2Se_3$. (c) Summary of the $V_{DC}$ dependence of the transverse resistance $R_H$ in four different devices. $R_H$ is defined as half of the difference between the saturation values of $dV_{24}/dI_{13}$ at large positive fields and negative fields.

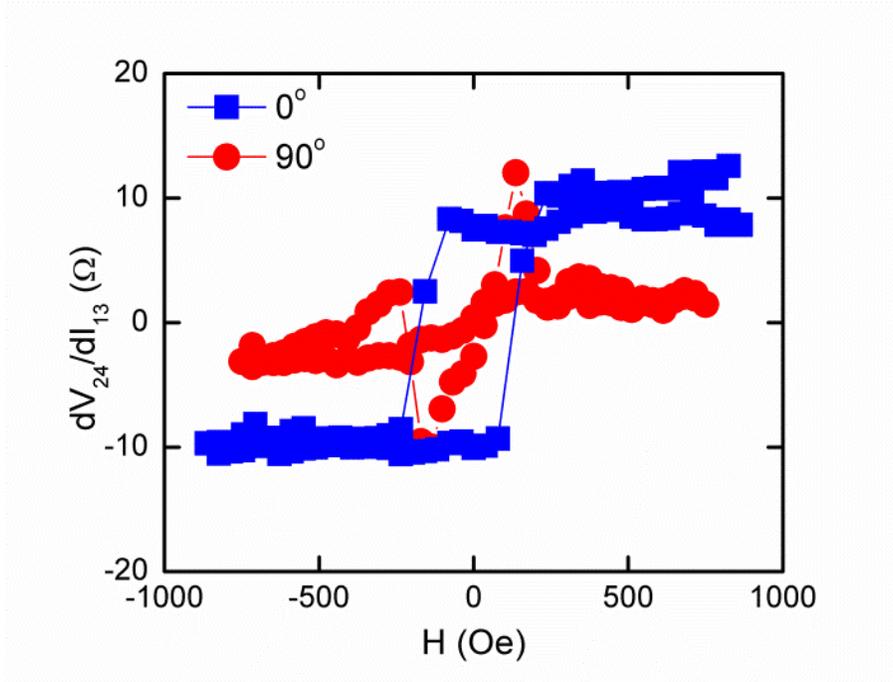

Figure 5. Field-dependent transverse differential resistance $dV_{24}/dI_{13}$ of $(Bi_{0.5}Sb_{0.5})_2Te_3$ in the tunneling configuration. The junction area is 10×3 μm$^2$. The magnetic field is swept along 0° (square) and 90° (circle) respectively. The dip and peak features at low fields of the 90° sweep come from the random selection of the magnetic moment orientation along the two degenerate states: 0° and 180°, as there is no symmetry breaking mechanism that would favor one direction over the other under a reasonably good field alignment.

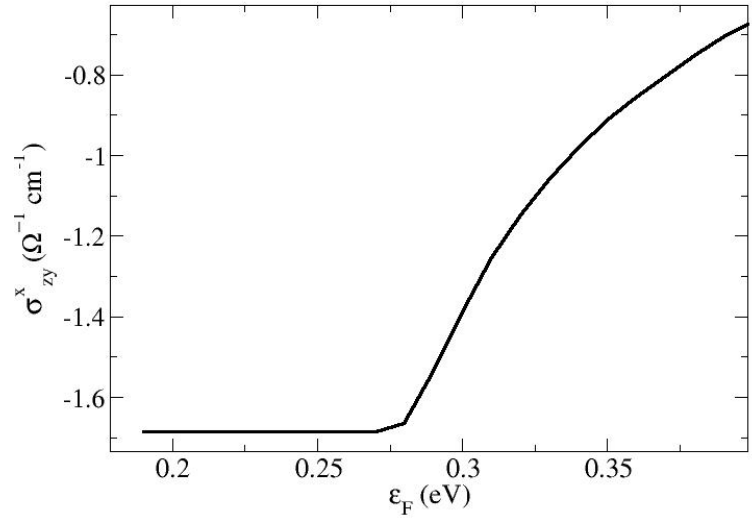

Figure 6. Calculated element of the bulk spin Hall conductivity tensor that contributes to the Hall resistance measured in our experiments. The temperature is taken to be 10K.

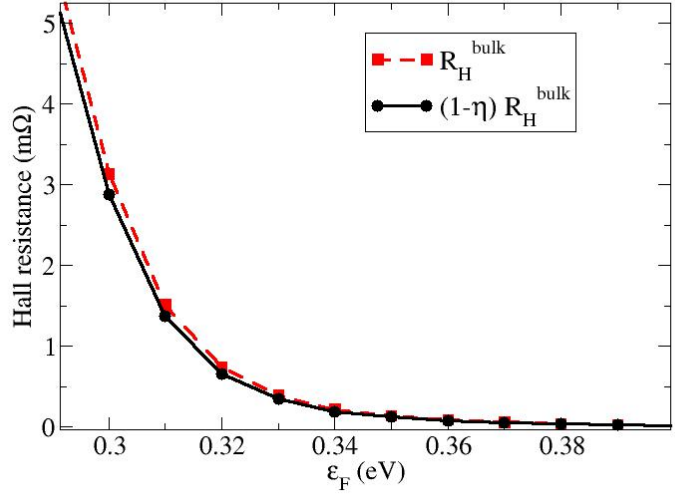

Figure 7: Calculated contribution from bulk states to the measured signal $R_H$ (Hall resistance) via the inverse spin Hall effect (Eq. (A.7)). The dashed (red) line assumes that all the injected current goes into the bulk. The solid (black) line accounts for the fact that only a fraction $1 - \eta$ of the injected current is carried by the bulk states. The factor $\eta$ is calculated as shown in Eq. (A.20).

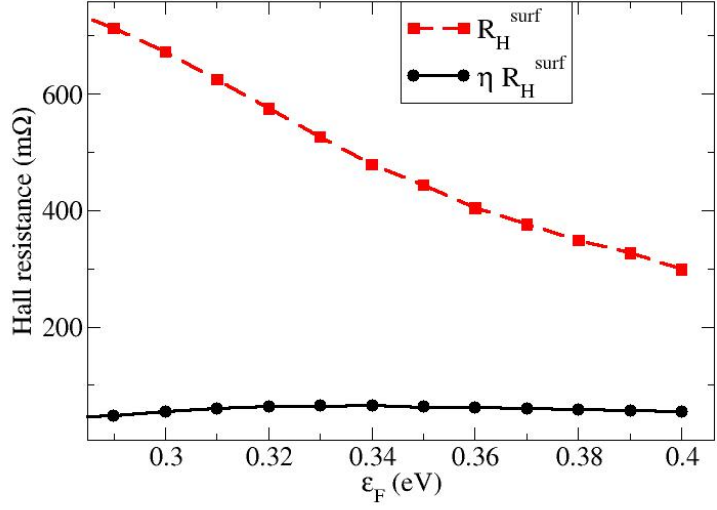

Figure 8: Calculated contribution from the surface states to the measured signal $R_H$ (Hall resistance), via the spin-galvanic effect (Eq. (A.9)). The dashed (red) line assumes that all the injected current goes onto the surface states. The solid (black) line accounts from the fact that only a fraction $\eta$ of the injected current is carried by the surface states.